# Impacts of demographic catastrophes

Ron W Nielsen aka Jan Nurzynski[1]

Environmental Futures Centre, Gold Coast Campus, Griffith University, Qld, 4222, Australia

November, 2013

Analysis of demographic catastrophes shows that, with the exception of perhaps only two critical events, they were too weak to influence the growth of human population. These results reinforce the conclusion that the concept of the Epoch of Malthusian Stagnation, the alleged first stage of growth claimed by the Demographic Transition Theory, is not supported by empirical evidence. They show that even if we assume that Malthusian positive checks are capable of suppressing the growth of population their impact was too weak to create the Epoch of Malthusian Stagnation.

**Introduction**[2]

In the previous publication (Nielsen aka Nurzynski, 2013a), we have analysed the effects of Malthusian positive checks. We have found that the *growth rate increases with the increasing intensity of positive checks*, the effect being opposite to the expected effect created by the mechanism of Malthusian stagnation. If the intensity of positive checks is small, the growth rate is also small but if the intensity of positive checks is large the growth rate is also large. Thus, empirical evidence indicates that Malthusian positive checks to not active the mechanism of Malthusian stagnation but that they activate the mechanism of Malthusian regeneration (Malthus, 1798; Nielsen aka Nurzynski, 2013b), the mechanism that acts against the forces of positive checks and repairs quickly any damage caused by these forces. Empirical evidence contradicts the fundamental concept of the Epoch of Malthusian Stagnation, the first stage proposed by the Demographic Transition Theory, suggesting that Malthusian regeneration mechanism restores the process of growth to its original trajectory and prevents any form of stagnation.

---

[1] r.nielsen@griffith.edu.au; ronwnielsen@gmail.com; http://home.iprimus.com.au/nielsens/ronnielsen.html

[2] Discussion presented in this publication is a part of a broader study, which is going to be described in a book (under preparation): *Population growth and economic progress explained*.

We have also pointed out that population data (Maddison, 2010; Manning, 2008; US Census Bureau, 2013) show that the growth rate in the distant past was small. Combining the two sources of information, the effect of positive checks expressed in increasing the growth rate and the evidence based on the population data showing that the growth rate was small, we have concluded (predicted) that *demographic crises in the distant past were too weak to influence the growth of human population*. We shall now check whether this prediction is confirmed by empirical evidence.

**Preliminary remarks**

Impacts of demographic catastrophes depend on their *intensity* (measured by the *death toll*), their *duration* and on the *size* of the affected population. However, even considering all these factors, impacts will not just depend on the number of people *directly* affected by a given crisis but more importantly on the number of people living in a certain country or a region where not all of the population was affected, and in the final analysis on the size of global population. It is, therefore, important to distinguish between local and global effects. Local death toll measured either in the number of killed people or as the percentage of the affected population may be high but it might not be high enough to affect significantly the growth of the world population or even the growth of regional population.

The death toll is usually given as the *number* of killed people and it is often impressively high. However, it is important to express it as the *percentage* of the population, because it is this number that can tell us whether demographic catastrophe was strong or weak, but even then, even if the percentage is high, it is important to distinguish between *local* effects and the effects embracing a significantly larger part of *global or regional* population.

The further back in time we extend our investigation the less we know about the demographic catastrophes, their intensity and their possible impacts, but we have enough information about the AD era to find out whether positive checks were strong enough to affect the growth of human population. In order to understand human population dynamics it is essential to identify the *main* driving force, or forces, of growth. The fundamental question is whether random forces of positive checks proposed by Malthus (1798) were strong enough to interact substantially with the major driving force, either to suppress growth or to stimulate the recovery mechanism. Even though we do not have much information about demographic catastrophes during the BC era we might expect that if they had a substantial effect locally, because the size of human population was small in any place, and consequently the percentage of killed people might have been large, their general impact was probably small because people lived in greater isolation.

It should be also noted that the recorded impacts of demographic catastrophes are likely to be exaggerated. Recorded death rates "are largest when the supporting evidence is skimpiest. When data are better, the death rates are usually lower and the percentage increases less" (Watkins & Menken, 1985, p. 651). For instance, both Durand (1960) and Fitzgerald (1936, 1947) claim that the impact of the An Lu-Shan Rebellion (AD 756-763) is probably exaggerated. Likewise, Russel (1968) and Twigg (1984) believe that the number of casualties caused by Justinian Plague (AD 541-542) is also grossly overestimated.

Another example is the Antonine Plague (AD 166-270), which was first estimated to have killed about 50% of the population of the Roman Empire (Seeck, 1910), but this estimate was later downgraded to 1-2%, or to the total number of casualties of 500,000-1,000,000 (Gilliam, 1961) and eventually upgraded to 7-10% or to a maximum of 5 million (Littman & Littman, 1973), the last estimate being still significantly smaller than the original estimate. It appears that the further back in time we go the larger is the possibility of exaggerated claims of the number of casualties.

Fortunately, we have a sufficiently large number of reliable accounts for the AD era to study the effects of demographic catastrophes. Furthermore, exaggerated estimates are not going affect our investigation because, while keeping in mind the possibility of the overestimated death tolls, our aim is not to be biased towards accepting only small estimates in order to prove that impacts of demographic catastrophes were small but to include any large estimates and investigate also their impacts. Contradicting indications based on exaggerated estimates of death tolls but producing small effects will represent a strong argument in favour of our prediction that impacts of demographic catastrophes were small.

We shall identify demographic catastrophes in the way they are reported in the literature. However, labelling them with just a single cause might not be accurate. For instance, a war considered as the main cause of a crisis might include famine but famine might be linked with pestilence. For example, during the Madras famine in the 1870s, about 40% of casualties were caused by smallpox and cholera (Lardinois, 1985). The Justinian Plague was accompanied by smallpox, diphtheria, cholera and influenza (Shrewsbury, 1970) and was "perhaps aided by wars, famines, floods and earthquakes" Scott and Duncan (2001, p. 5). Likewise, "a number of epidemics in France were preceded by famine, sometimes in conjunction with bad weather conditions" (Scott & Duncan, 2001, p. 105) whereas "frequent and virulent outbreaks in France during 1520-1600 were accompanied by food shortages, famines, flooding, peasant uprisings and religious wars" (Scott & Duncan, 2001, p. 291).

While drawing from primary sources about the frequency and intensity of demographic catastrophes, the presented survey has been also assisted by some useful compilations (Austin Alchon, 2003; Kohn, 1995; Spignesi, 2002; White, 2011).

**Survey of demographic catastrophes**

One of the earliest recorded plagues was the Asiatic disease identified now as tularaemia, a bacterial disease caused by *Francisella tularensis*, first transported to Egypt around the turn of the 18th century BC from the Middle East by contaminated ships (Trevisanato, 2004, 2007). The same disease has been also probably recorded in the Bible as causing a great number of deaths among Philistines in the city of Ashdot, the event dated either to around 1000 BC (Khan, 2004) or to 1320 BC (Cunha & Cunha, 2006).

Early recorded plagues include also a viral haemorrhagic fever in Egypt between 1500-1350 BC (Duncan & Scott, 2005) but it might have been the same disease as recorded earlier in Egypt and the same plague that decimated Philistines. Incidentally, Duncan and Scott (2005) claim that the Black Death was not a bubonic plague caused by bacterium *Yersinia pestis*, as traditionally claimed, but rather that is was a viral haemorrhagic fever, which according to them includes also the plagues of Mesopotamia (700-350 BC), the Plague of Athens (430-427 BC), the Plague of Justinian (AD 541-542), Plagues of Islam (AD 627-744), plagues in Asia minor (1345-1348), and the plague of Denmark and Sweden (1710-1711).

The epidemic of Athens (460-399 BC) is claimed to have killed 25% of Athenian army and a great number of civilians (Austin Alchon, 2003). It created a turning point in the history of Greece (Ross, 2008). However, it is also claimed that this plague killed 50% of the army of Pericles and 50% of the navy coming to the rescue from the Piraieus (Beran, 2008). The plague was triggered by overcrowding of Athens when Spartan's attacks prompted rural population to seek shelter in that city, which was already housing a relatively large number of people, an estimated 300,000 citizens and around 3 million slaves.

The earliest and the most lethal demographic catastrophes in the AD era was probably the Red Eyebrows Revolt commencing around AD 2. Durand (1960) lists the Chinese population as being 59.6 million in AD 2 and 21 million in AD 57, a total decline of 38.6 million in 55 years, but according to him this decline was only partly caused by this rebellion (Durand, 1960p. 216). However, he then discusses possible inaccuracies in these estimates and presents corrected numbers as being 74 million in AD 2 and 45 million in AD 88, for the entire Chinese Empire, a drop of 29 million in 86 years. He also estimates 71 million and 43 million, respectively, for the China proper, representing a total drop of 28 million (Durand, 1960, p. 221). By China proper he means the current 18 provinces. He uses this estimate in his graph (Durand, 1960, p. 247).

Durand points out that estimates of the size of the population at the time of demographic catastrophes might be inaccurate. "Even if such huge loss were conceivable, it would be naïve to suppose that accurate count of the survivors could have been carried out in the midst of the ensuing chaos" (Durand, 1960, p. 224). White (2011) attributes only 10 million of casualties to the Red Eyebrows Revolt. However, to estimate the maximum impact we shall accept the death toll of 29 million, and we shall attribute it entirely to the Red Eyebrows Revolt, keeping in mind that this figure might have been overestimated by the factor of three.

The same applies to the An Lu-Shan Rebellion (AD 756-763). Records appear to show the death toll of 36 million but White (2011) attributes only 13 million. Here again, to lean in favour of the concept of Malthusian Stagnation we shall assume 36 million casualties as suggested by the census rather than 13 million as assumed by White (2011).

> Between A.D. 705 and 755 to all appearances the census machinery functioned much more effectively; but after 755 it broke down again. The recorded number of persons dropped from nearly 53 millions in the year 755 to only 17 millions in 760. During this time China was torn by revolts which were suppressed with bloody force, including the notorious rebellion of An Lu-Shan. Many historians have affirmed that 36 million lives were lost as a result of these violent events, but Fitzgerald and others have shown that this is incredible (Durand, 1960, p. 223; Fitzgerald, 1936, 1947).

The effect of the Plague of Justinian is hard to estimate because of confusing claims. The plague is claimed to have reduced the population of Constantinople by 40% between AD 541 and 542 (Austin Alchon, 2003). However, Cunha and Cunha (2006) claim AD 542-590 for its duration and a 30% reduction of the population of the Roman Empire, or a maximum of about 14 million out of the total of 48 million (Maddison, 2006; Seeck, 1921). "The plague so weakened the Roman Empire that not long after the plague had passed, Roman borders were overrun by Huns, Goths, Moors, and other 'barbarians'" (Cunha & Cunha, 2006).

It is claimed that in AD 549 the plague emerged in Britain (Carmichael, 2009) and in AD 610, assisted by the Eurasian Silk Road, it spread to China (Ross, 2008) continuing its devastation until around AD 700 (Duncan & Scott, 2005) and killing probably a maximum of about 100 million people but over an unspecified time (Ross, 2008), making it difficult to calculate its impact. It is also claimed that "The Plague of Justinian recurred in discernible cycles of about nine to twelve years" (Dols, 1974, p. 373). In order to estimate the maximum impact we shall assume that the total number of casualties was 25 million in a short time of only between AD 541 and 542 (Rosen, 2007). Had we used the suggestion of Cunha and Cunha (2006) the estimated impact would have been much smaller because the number of claimed casualties was small and the death toll was spread over a much longer time.

The Black Death (1343-1351) is claimed to have killed over 60% of the urban population in Asia, about 30% of the population of the Middle East and 30-60% of the population of Europe (Hawas, 2008). Beran (2008) claims that in many cities the death toll was over 90%, creating incredible hardship for the surviving few, adding perhaps to the death toll caused by the lack of food and safe shelter, let alone by complications created by decaying corpses. About 20% of the population of England died between AD 1348 and 1350 and a total of 50% by AD 1400 (Gilliam, 1961). Depending on the affected area, mortality rates varied between 25% and 70% (Cunha & Cunha, 2006). The Black Death appears to have been the greatest demographic catastrophe ever recorded.

If plagues were not powerful enough to kill large number of people, they were also used as biological weapons by employing a gruesome practice of catapulting infected corpses at the walls of fortifications or hurling them over the walls using trebuchets. This ghastly method was used by Greeks, Romans and other attackers between 300 BC and AD 1100, and by Tartars in 1346 against the residents of Genoa (Cunha & Cunha, 2006; Khan, 2004).

Other examples of large local casualties caused by demographic catastrophes include smallpox in Japan (AD 812-814), killing about half of the population of that country (Austin Alchon, 2003); the 1696 famine killing between 25% and 30% of the population of Finland (Jutikkala, 1955); the 1770 famine in Bengal, killing about 30% of the population (or a total of 10 million) and the 1376 famine in Italy, killing 60% of the population (Ghose 2002; Keys, Brozek, Henschel, Mickelsen & Taylor, 1950; Walford, 1878).

According to Mallory (1926), between AD 620 and 1619, 18 provinces of China experienced 1015 draughts, or about one per year. However, they were unevenly distributed, illustrating that while the number of casualties and impacts of demographic catastrophes might be high in a small and strongly-affected regions, the effects could be much less severe when averaged over a larger number of human population.

There was a total of 443 draughts in the Northern Division, 352 in the Central Division and 220 in the Southern Division. However, even within the same division, the number of draughts varied significantly between various districts. For instance, in the Northern Division, Honan District experienced a total of 112 draughts but Kansu Division only 4. In the Central Division, the largest number of draughts (113) was in the Chekiang District and the smallest (28) in the Anhwei District. In the Southern Division the number of draughts varied between 4 and 59 per district.

The list of significant lethal events in China includes: 60-70% of troops killed during a single military engagement in AD 16; 70% of Mongolians killed by hunger in AD 46; 30-40% of troops killed in AD 162; about 70% of troops killed in a single military engagement and by famine and epidemic; close to 100% killed by locusts and famine in AD 312 in the northern and central China; over 30% killed in Shantung in AD 762; over 50% in Chekiang in AD 806; 30-40% in Hupeh, Kinagsu and Anhui in AD 891; 90% in Hopei in 1331; 50% of troops between 1351-1352; over 70% in Shansi in AD 135; 60-70% in Hupeh in 1354, and 100% in various towns and villages in Hunan in 1484 (McNeill, 1976; see also Austin Alchon, 2003)

It is claimed that in Mexico, 25-50% of the population died of smallpox (1520-1521), 60-90% probably of typhus (1531-1532), and over 50% of either the bubonic plague or typhus between 1576 and 1581 (Austin Alchon, 2003; Motolinía aka Fray Toribio de Benavente o Motolinía, 1971; del Paso y Troncoso, 1940; Prem, 1992).

The estimated death toll in the Andes between 1524 and 1591 include 30-50% of smallpox (1524-1527), 25-30% of measles or bubonic plague (1531-1533), 15-20% of influenza, measles and smallpox (1558-1559), and about 50% of influenza, measles, smallpox and typhus between 1585 and 1591 (Cook, 1981; Dobyns, 1963). Dobyns (1993) gives also many examples of large death tolls, sometimes as high as 98% but most often close to 80-90%, caused by diseases among Native American population.

So, it appears that humans always lived with the threats and deadly effects of demographic catastrophes strong enough to reduce often substantially the size of local populations, the severity and frequency of these events appearing to be in support of the Demographic Transition Theory and of the idea of the prolonged Epoch of Malthusian Stagnation, if we still want to insist that the mechanism of Malthusian stagnation was active. We have already shown that this mechanism is not active now and probably was never active because empirical evidence suggests that positive checks activate the efficient Malthusian regeneration mechanism (Malthus, 1798; Nielsen aka Nurzynski, 2013a, 2013b).

However, for the sake of argument, let us assume that the mechanism of Malthusian stagnation was active in the past. The question now is whether demographic crises were strong enough to affect the growth of human population. More specifically, if we still want to assume that positive checks activate the Malthusian stagnation mechanism rather than the Malthusian replacement mechanism, the question is whether these apparently strong and frequent critical events occurring in various parts of the world had a significantly strong influence on the growth of global population, the influence, which might have been reflected in stagnation, Malthusian oscillations and fluctuations, or whether the effects were not strong enough to interfere with a perhaps much stronger and dominating driving force of growth. The question is also whether these opposing forces of critical events were even strong enough to influence the regional growth of the population over a significantly long time. This last question is not investigated in this article but is addresses in the forthcoming book mentioned in the footnote to the *Introduction*.

**Impacts of demographic catastrophes**

Results of the survey of demographic catastrophes are presented in Table 1. The Epoch of Malthusian Stagnation is claimed to have "persisted until the end of the 18th century" (Galor, 2005a, p. 180). Consequently, the results of the survey extend to around 1900 to allow for about 100-year overlap between the alleged end of the Epoch of Malthusian Stagnation and the claimed new regime of growth. The BC era is not included because we do not have enough reliable information about that time, but the survey for the AD era should be sufficient to assess the reliability of the concept of the Epoch of Malthusian Stagnation.

In this table, impacts of demographic catastrophes are measured by (1) the *number* of killed people, (2) the *duration* of crisis, (3) the *total percentage* of the population killed during the entire time of crisis (total impact), and (4) by the *percentage* of the population killed *per year*. Other gauge indicators can be also introduced. They are discussed in the forthcoming book.

The most remarkable and unexpected result of this survey is that impacts of demographic crises were in general surprisingly small. Total impacts of up to around 10% can be ignored because such occasional deviations are too small to be reflected in the growth trajectory. Even occasional displacement as large as 20% is unlikely to affect significantly the overall trend. Our survey shows that 84% of the *major* demographic catastrophes were characterised by the total impact of less than 5%, and 48% by the total impact of less than 1%. It should be also mentioned that Table 1 lists only large catastrophes associated with total death tolls of 1,000,000 or larger. Had we listed all reported catastrophes, the percentage of demographic crises having negligible impacts on the growth of human population would have been even larger, or alternatively the percentage of significant demographic crises would have been negligibly small.

**Table 1.** Demographic catastrophes and their impacts

| Event | Onset (AD) | Death Toll | Duration [Years] | Total Impact [%] | Impact pa [%] |
|---|---|---|---|---|---|
| Red Eyebrows Revolt | 2 | 29,000,000 | 87 | 11.5 | 0.13 |
| Antonine Plague | 166 | 5,000,000 | 15 | 2.2 | 0.15 |
| Plague of Justinian | 541 | 25,000,000 | 2 | 12.5 | **6.23** |
| An Lu-Shan Rebellion | 756 | 36,000,000 | 8 | **15.4** | 1.93 |
| N. Egypt Earthquake | 1201 | 1,500,000 | 1 | 0.5 | 0.46 |
| Mongolian Conquest | 1260 | 40,000,000 | 35 | 11.3 | 0.32 |
| Great European Famine | 1315 | 7,500,000 | 3 | 2.0 | 0.68 |
| Famine in China | 1333 | 9,000,000 | 15 | 2.4 | 0.16 |
| Black Death | 1343 | 75,000,000, | 9 | **19.7** | 2.19 |
| Fall of the Yuan Dynasty | 1351 | 7,500,000 | 18 | 1.9 | 0.11 |
| Sweating Sickness | 1485 | 3,000,000 | 67 | 0.6 | 0.01 |
| Mexico Smallpox Epidemic | 1520 | 4,000,000 | 2 | 0.8 | 0.42 |
| French Wars of Religion | 1562 | 3,000,000 | 37 | 0.6 | 0.02 |
| Russia's Time of Trouble | 1598 | 5,000,000 | 16 | 0.9 | 0.06 |
| Fall of the Ming Dynasty | 1618 | 25,000,000 | 27 | 4.3 | 0.16 |
| Thirty Years War | 1618 | 7,000,000 | 31 | 1.2 | 0.04 |
| Deccan Famine in India | 1630 | 2,000,000 | 2 | 0.3 | 0.17 |
| Famine in France | 1693 | 2,000,000 | 2 | 0.3 | 0.15 |
| Bengal Famine | 1769 | 10,000,000 | 5 | 1.2 | 0.23 |
| Napoleonic Wars | 1803 | 4,000,000 | 13 | 0.4 | 0.03 |
| Famines in China | 1810 | 22,500,000 | 2 | 2.3 | 1.13 |
| Great Irish Famine | 1845 | 1,000,000 | 6 | 0.1 | 0.01 |
| Famine in China | 1846 | 11,300,000 | 1 | 1.0 | 0.96 |
| Taiping Rebellion | 1850 | 20,000,000 | 15 | 1.6 | 0.11 |
| Famine in India | 1866 | 1,000,000 | 1 | 0.1 | 0.08 |
| Famine in Rajputana | 1869 | 1,500,000 | 1 | 0.1 | 0.11 |
| Famine in Persia | 1870 | 2,000,000 | 2 | 0.1 | 0.07 |
| Famine in N. China | 1876 | 13,000,000 | 3 | 0.9 | 0.31 |
| British India Famine | 1876 | 17,000,000 | 25 | 1.1 | 0.05 |
| Yellow River Flood | 1887 | 2,000,000 | 1 | 0.1 | 0.13 |
| Famine in India | 1896 | 8,300,000 | 6 | 0.5 | 0.08 |

There were *only two* catastrophes that might have caused a temporary distortion in the growth of human population: the An Lu-Shan Rebellion with its maximum estimated total impact of 15.4% and the Black Death with its estimated impact of 19.7%. *Two outstanding but moderate demographic catastrophes could not have had a strong influence on the growth of human population* even if we assume that the mechanism of Malthusian was active. If, however, as suggested by empirical evidence, the Malthusian regeneration mechanism was active, even large losses would have been quickly replaced and the damage repaired. We should also remember that we have listed the estimated maximum impacts. The actual impacts could have been smaller.

In estimating impacts of demographic crises, we should also consider overlapping or closely-spaced catastrophes. A single catastrophe might be too weak to cause any significant damage, but if combined with closely-spaced catastrophes, their *collective impact* could be large. The survey shows *only one* such significant cluster of events: Mongolian Conquest with the total estimated death toll of 40 million; the Great European Famine, 7.5 million; the 15-year Famine in China, 9 million; Black Death, 25 million; and the Fall of Yuan Dynasty, 7.5 million. However, Black Death has been already identified as possibly having a noticeable effect, which means again that out of all catastrophic events *only two* might have been strong enough to cause any tangible distortion in the population trajectory: the An Lu-Shan Rebellion and the combination of five closely-spaced catastrophes. All other catastrophes had negligible effect on the growth of human population.

**Summary and conclusions**

One might expect that many demographic catastrophes were not duly recorded, but the results of this survey show that the *overwhelming majority of recorded catastrophic events were too* *weak to have any effect on the growth of human population.* This conclusion contradicts the postulate of the existence of the Epoch of Malthusian Stagnation, the first stage of growth proposed by the Demographic Transition Theory, even if, contrary to the empirical evidence, we still assume that the mechanism of stagnation rather than the mechanism of regeneration was active.

As pointed out earlier (Nielsen aka Nurzynski, 2013b) the mechanism of regeneration was first suggested by Malthus (1978) and confirmed empirically by the study of the dependence of the growth rate on the level of deprivation (Nielsen aka Nurzynski, 2013a), the study based on one of many sets of data, which were not available to Malthus. Considering the strongly limited information available to him, he must have been endowed with the exceptional power of observation and foresight, surpassing the power of observation of many who tried to follow his footsteps and yet failed to notice the multitude of empirical evidence pointing to the correct understanding of the human population dynamics.

Results of this survey show that demographic crises were so weak that even if they activated the Malthusian regeneration mechanism, any damage caused by them was quickly and efficiently repaired, which leads us to yet another conclusion (expectation) that *global or even perhaps regional trajectories of growth of human population will not show any major disturbances*. Population growth trajectories should display robust and stable characteristics.

The major points of our investigation can be summarised as follows:

1. Empirical evidence (Nielsen aka Nurzynski, 2013a) suggests that positive checks do not suppress the growth of human population but activate an efficient *regeneration mechanism* (Malthus, 1798; Nielsen aka Nurzynski, 2013b).
2. There was only one combination of five closely-spaced demographic catastrophes (Mongolian Conquest, Great European Famine, the 15-year Famine in China, Black

Death, and the Fall of Yuan Dynasty) that may have had a small effect on the growth of human population. Another possibly strong effect might have been produced by the An Lu-Shan Rebellion, but the death toll associated with this event is almost certainly overestimated (Durand, 1960; Fitzgerald, 1936, 1947).

3. Demographic catastrophes (positive checks) were too weak to alter the course of the growth of human population.

4. Demographic crises were so weak that the Malthusian regeneration mechanism, if activated, was able to repair any damage quickly and efficiently.

5. The growth of human population must have been controlled by a yet to be identified strong driving force.

Jutikkala, E. (1955). The great Finish famine 1696-97. *Scand. Econ. Hist. Rev., 3*, 48-63.

Keys, A., Brozek, J., Henschel, A., Mickelsen, O., & Taylor, H. L. (1950). *The biology of human starvation*. Minneapolis: University of Minnesota Press.

Khan, I. A. (2004). Plague: the dreadful visitation occupying the human mind for centuries. *Transactions of the Royal Society of Tropical Medicine and Hygiene*, *98*, 270-277.

Kohn, G. C. (1995). *The Wordsworth encyclopedia of plague and pestilence.* New York: Facts on File.

Lardinois, R. (1985). Famine, epidemics and mortality in South India: A reappraisal of the demographic crisis of 1876-78. *Economic and Political Weekly*, *20*(11), 454- 465.

Littman, R. J., & Littman, M. L. (1973). Galen and the Antonine Plague. *The American Journal of Philology, 94*(1), 243-255.

Maddison, A. (2006). *The World Economy.* Paris: OECD.

Maddison, A. (2010). *Historical statistics of the world economy: 1-2008 AD*. http://www.ggdc.net/maddison/Historical Statistics/horizontal-file_02-2010.xls

Mallory, W. H. (1926). *China: Land of famine*. New York: American Geographical Society.

Malthus, T. R. (1798). *An Essay on the Principle of Population as It Affects the Future Improvement of Society, with Remarks on the Speculations of Mr Godwin, M. Condorcet, and Other Writers*. London: J. Johnson.

Manning, S. (2008). *Year-by-Year World Population Estimates: 10,000 B.C. to 2007 A.D*. http://www.digitalsurvivors.com/archives/worldpopulation.php and references therein.

McNeill, W. H. (1976). *Plagues and people*. Garden City, NY: Anchor Books.
17